\documentclass[aps,pra,twocolumn,10pt]{revtex4-1}

\usepackage{natbib}
\usepackage{amsmath,amssymb,bm}
\usepackage{graphics}
\usepackage{color}

\newcommand{\Dcirc}{\mathcal{D}^{\circ}}

\begin{document}

\title{Quantifying entanglement of parametric down-converted states \\ in all degrees of freedom}

\author{Filippus S. Roux}
\email{froux@nmisa.org}
\affiliation{National Metrology Institute of South Africa, Meiring Naud{\'e} Road, Brummeria, Pretoria 0040, South Africa}

\begin{abstract}
The amount of entanglement that exists in a parametric down-converted state is investigated in terms of all the degrees of freedom of the state. We quantify the amount of entanglement by the Schmidt number of the state, represented as a pure bipartite state by tagging the down-converted photons in terms of orthogonal states of polarization with the aid of type II phase-matching. To facilitate our calculations, we use a Wigner functional approach, which allows the incorporation of the full infinite dimensional spatiotemporal degrees of freedom. A quantitative example with reasonably achievable experimental conditions is considered to demonstrate that extremely large Schmidt numbers are achievable.
\end{abstract}

\maketitle


\section{\label{intro}Introduction}

Parametric down-conversion \cite{mandelspdc} is a nonlinear optical process that is widely used in the preparation of entangled quantum states for photonic quantum information systems. It is used to produce correlated photons for remote clock synchronization \cite{homsync2,homsync1}, in quantum ghost imaging \cite{ghostph}, in quantum teleportation \cite{advtelport}, in protocols for quantum key distribution \cite{scarani}, and it is also used to prepare squeezed states for multi-photon applications \cite{sqlight,squreview}, including their use for sub-shot-noise observations of gravitational waves \cite{ligosqu}. While some applications focus on the spatiotemporal degrees of freedom of individual photon pairs in the parametric down-converted state (PDCS), others harness its multi-photon properties with less emphasis on the spatiotemporal degrees of freedom.

Higher dimensional states allow more entanglement; they provide more information capacity and more security in quantum cryptology \cite{bechmann,cerf}. For a pure state, one can quantify the dimensionality of the entangled state by the Schmidt number --- the average number of modes in the Schmidt decomposition of the state. Orbital angular momentum modes \cite{allen,mtt} are often used to represent the PDCS as a high-dimensional state with a relatively large model Schmidt number \cite{law,pires,miatto1,gattispdc}. But it captures only the spatial degrees of freedom.

The particle-number degrees of freedom in the PDCS are also entangled, with the corresponding Schmidt basis being the Fock basis. One can expect that the total entanglement in terms of all the degrees of freedom in the PDCS would be much higher than estimates solely based on the spatiotemporal degrees of freedom in the state would indicate. However, the calculation of such a full Schmidt number is severely challenging, as evident from recent attempts \cite{adesso,leuchs1,bsqvac, sharapova,namiki}. These studies impose restrictions on the degrees of freedom, for example by considering only a finite number of discrete modes.

Here, we determine the entanglement of the PDCS in terms of the full Schmidt number, including the infinite-dimensional spatiotemporal and particle-number degrees of freedom. This calculation is facilitated by a Wigner functional formalism \cite{stquad,ipfe}. We apply the semi-classical approximation where the pump is assumed not to be significantly affected by the process. Thus, the parametric down-conversion process becomes equivalent to a squeezing operation on the vacuum state. Being interested in the maximal potential entanglement in the state prior to any processing or measurement, we'll assume that the PDCS is a pure state, ignoring any loss occuring during the preparation stage. At some point, we'll assume some minor processing of the state for the sake of tractability, but such processing is assumed not to affect the purity of the state.

To compute the Schmidt number, the PDCS must be represented as a bipartite state. The photons in a pure PDCS are all produced in pairs, but they are not necessarily distinguishable, unless there is some mechanism to tag them. Often, some form of post-selection is used to tag the photons. In practice, such post-selection can be done by using two complimentary displaced apertures or different complimentary wavelength filters. Here, we'll use type II phase-matching, which tags the two photons with different states of polarization. The resulting {\em bipartite squeezed state} is analogues to a two-mode squeezed state, where the two ``modes'' are the two states of polarization and does not restrict the other degrees of freedom.

The paper is organized as follows. We generalize the bipartite squeezing operator to incorporate all the spatiotemporal degrees of freedom in Sec.~\ref{squop}. It is used in Sec.~\ref{vol} to derive an expression for the full Schmidt number with the aid of a Wigner functional approach. In Sec.~\ref{pdcp}, the detailed kernel for type II parametric down-conversion is incorporated into the expression of the expression of the full Schmidt number. This expression consists of a functional determinant, which is evaluated in Sec.~\ref{calc} by applying suitable approximations. By considering typical experimental conditions, we present numerical curves for the full Schmidt number in Sec.~\ref{numexp}, showing a dramatic increase in the entanglement of the PDCS. In Sec.~\ref{concl}, we end with some conclusions.

\section{\label{squop}Squeezing operator}

Including all the spatiotemporal degrees of freedom in the bipartite squeezing operator, one can express it as
\begin{equation}
\hat{S} = \exp\left(\tfrac{1}{2}\hat{a}_e\diamond\zeta_{\text{eo}}^*\diamond\hat{a}_o
-\tfrac{1}{2}\hat{a}_e^{\dag}\diamond\zeta_{\text{eo}}\diamond\hat{a}_o^{\dag}\right) ,
\label{gensqu}
\end{equation}
where the squeezing parameter becomes a kernel function and the subscripts $e$ and $o$ represent the extra-ordinary and ordinary states of polarization, respectively, as produced by parametric down-conversion with type II phase-matching. The contractions of the ladder operators on the kernels are represented by the notation
\begin{align}
\hat{a}_e\diamond\zeta_{\text{eo}}^*\diamond\hat{a}_o \equiv & \int \hat{a}_e(\mathbf{k}_1)\zeta_{\text{eo}}^*(\mathbf{k}_1,\mathbf{k}_2)\hat{a}_o(\mathbf{k}_2) \nonumber \\
& \times \frac{\text{d}^3 k_1}{(2\pi)^3\omega_1}\ \frac{\text{d}^3 k_2}{(2\pi)^3\omega_2} .
\label{defafbind}
\end{align}
In terms of this notation, the Wigner functional for a bipartite squeezed vacuum state reads
\begin{align}
W_{\text{bsv}}[\alpha,\beta] = & \mathcal{N}_0^2 \exp\left[-2\left(\alpha^*\diamond\mathcal{C}\diamond\alpha
+ \beta^*\diamond\mathcal{C}\diamond\beta \right. \right. \nonumber \\
& \left. \left. + \alpha^*\diamond\mathcal{S}\diamond\beta^* + \alpha\diamond\mathcal{S}^*\diamond\beta\right)\right] ,
\label{wigpure2vak}
\end{align}
where $\alpha$ and $\beta$ are the fields for the two different states of polarization, $\mathcal{N}_0$ is a normalization constant, and
\begin{align}
\begin{split}
\mathcal{C} \equiv & \cosh_{\diamond}\left(2|\zeta_{\text{eo}}|\right) \\
\mathcal{S} \equiv & \exp_{\diamond}(i\theta_{\text{eo}})\diamond\sinh_{\diamond}\left(2|\zeta_{\text{eo}}|\right) ,
\end{split}
\label{defcs}
\end{align}
with $\zeta_{\text{eo}}=|\zeta_{\text{eo}}|\diamond\exp(i\theta_{\text{eo}})$ \cite{Note1}. The subscript $\diamond$ indicates that the products in the expansion of a function are $\diamond$-contractions and the first term is a Dirac delta function.

\section{\label{vol}Full Schmidt number}

In order to quantify the entanglement in the pure bipartite squeezed vacuum state, we use a Wigner functional approach \cite{stquad,ipfe} to compute the full Schmidt number. The Wigner functional approach alleviates this calculation significantly. We start by computing the partial trace of Eq.~(\ref{wigpure2vak}) over $\beta$, which is done by performing the functional integration over $\beta$. In the process, we use the fact that $\mathcal{C}$ and $\mathcal{S}$ commute because they involve the same kernel. The result is
\begin{align}
W_\text{pt}[\alpha] = & \int W_{\text{bsv}}[\alpha,\beta]\ \Dcirc[\beta] \nonumber \\
 = & \frac{\mathcal{N}_0}{\det\{\mathcal{C}\}} \exp\left(-2\alpha^*\diamond\mathcal{C}^{-1}\diamond\alpha\right) .
\end{align}
The Wigner functional of the partial trace is that of a mixed state and has the form of a thermal state. The inverse of the full Schmidt number is given by the purity of the partial trace:
\begin{equation}
\frac{1}{\mathcal{K}} = \int W_\text{pt}^2[\alpha]\ \Dcirc[\alpha] = \frac{1}{\det\{\mathcal{C}\}} .
\end{equation}
Hence, the full Schmidt number can be expressed as
\begin{equation}
\mathcal{K} = \det\{\mathcal{C}\}
\equiv \exp\left(\text{tr}\left\{\ln_{\diamond}\left[\cosh_{\diamond}\left(2|\zeta_{\text{eo}}|\right)\right]\right\}\right) .
\label{sndef}
\end{equation}
Up till now, no approximations were employed, apart from the semi-classical approximation. The representation of the full Schmidt number in terms of a cosh-function is expected from the equivalent expressions based on the particle-number degrees of freedom only. The functional nature of the cosh-function and the determinant shows that the spatiotemporal degrees of freedom are also incorporated here.

\section{\label{pdcp}Parametric down-conversion}

For a quantitative evaluation of Eq.~(\ref{sndef}), we need an expression for the parametric down-conversion kernel. In addition to the phase-matching conditions, there are other experimental details that affect the form of this kernel. Here, we'll assume degenerate collinear operation. The restrictions thus imposed already reduces the amount of entanglement in the state. However, we pay the prize for the sake of definitiveness and tractability.

The unitary operator for parametric down-conversion with type-II phase-matching is
\begin{equation}
\hat{U}_{\text{pdc2}} \equiv \exp\left( \hat{a}_e\diamond H_{\text{eo}}^*\diamond\hat{a}_o - \hat{a}_e^{\dag}\diamond H_{\text{eo}}\diamond\hat{a}_o^{\dag}\right) ,
\label{ueosv}
\end{equation}
where
\begin{align}
H_{\text{eo}}(\mathbf{a}_1,\mathbf{a}_2) = & \eta\ \text{sinc}\left(\tfrac{1}{2}\pi^2w_{\text{p}}^2\beta
|\mathbf{a}_1-\mathbf{a}_2|^2\right) \nonumber \\
 & \times \exp\left(-\pi^2 w_{\text{p}}^2|\nu_1\mathbf{a}_1+\nu_2\mathbf{a}_2|^2\right) ,
\label{spdckern2}
\end{align}
with $\mathbf{a}$ being the two-dimensional transverse spatial frequency vector, $\nu_1=n_o/n_3$, $\nu_2=n_{\text{eff}}/n_3$, and
\begin{align}
\begin{split}
\eta & = \frac{\pi^2\alpha_0\sigma_{\text{II}}L w_{\text{p}}}{n_3^2 \lambda_{\text{p}}^2} \sqrt{\frac{\pi\delta\lambda}{2\lambda_{\text{p}}}} , \\
\beta & = \frac{n_o n_{\text{eff}} L\lambda_{\text{p}}}{\pi n_3 w_{\text{p}}^2} .
\end{split}
\label{pardefs}
\end{align}
Here, $\alpha_0$ is the complex amplitude of the coherent state representing the pump (the only complex quantity in the expression and not to be confused with the field $\alpha$ in the expression of the Wigner functional), $\sigma_{\text{II}}$ is the nonlinear coefficient of the crystal for type-II phase-matching, expressed as a cross-section (with units of area), $L$ is the length of the nonlinear crystal, $w_{\text{p}}$ is the waist radius of the pump beam, $\delta\lambda$ is the wavelength bandwidth of the pump, $n_o$ is the ordinary refractive index, $n_{\text{eff}}$ is the effective extra-ordinary refractive index, $n_3=\tfrac{1}{2}(n_o+n_{\text{eff}})$ and $\lambda_{\text{p}}$ is the pump wavelength. The nonlinear coefficient also depends on the wave vectors of the three beams, but for the conditions considered here, these dependences are expected to be weak enough to ignore.

Comparing Eq.~(\ref{ueosv}) to Eq.~(\ref{gensqu}), we see that one can substitute $|\zeta_{\text{eo}}|\rightarrow 2|H_{\text{eo}}|$ into Eq.~(\ref{sndef}), to obtain
\begin{align}
\mathcal{K} = & \det\left\{\cosh_{\diamond}\left[4|\eta| \left|\text{sinc}\left(\tfrac{1}{2}\pi^2 w_{\text{p}}^2\beta |\mathbf{a}_1-\mathbf{a}_2|^2\right)\right| \right. \right. \nonumber \\
 & \left. \left. \times \exp\left(-\pi^2 w_{\text{p}}^2|\nu_1\mathbf{a}_1+\nu_2\mathbf{a}_2|^2\right)\right]\right\} .
 \label{hoof1}
\end{align}

\section{\label{calc}Calculation}

In its most general form, the functional determinant of a functional cosh-function is intractable. However, there are ways to approach the problem. Our aim here is not an exhaustive analysis for all possible conditions, but a representative example to see what is possible. For example, if $|\eta|$ is very small (close to zero), then we have
\begin{equation}
\cosh_{\diamond}\left(4|H_{\text{eo}}|\right) \approx \mathbf{1}+8|H_{\text{oe}}|\diamond|H_{\text{eo}}| ,
\end{equation}
where $\mathbf{1}$ denotes a Dirac delta function. Then, the determinant becomes
\begin{equation}
\mathcal{K} \approx \exp\left(8\text{tr}\left\{|H_{\text{oe}}|\diamond|H_{\text{eo}}|\right\}\right) .
\label{trh2}
\end{equation}
This trace can be calculated to give
\begin{equation}
8\text{tr}\left\{|H_{\text{oe}}|\diamond|H_{\text{eo}}|\right\} = \frac{|\eta|^2}{\pi w_{\text{p}}^4\beta} .
\label{trh2a}
\end{equation}
In terms of Eq.~(\ref{pardefs}), the full Schmidt number becomes
\begin{equation}
\mathcal{K} \approx \exp\left(\frac{|\eta|^2}{\pi w_{\text{p}}^4 \beta}\right)
= \exp\left(\frac{\pi^5 |\alpha_0|^2\sigma_{\text{II}}^2 L \delta\lambda}{2 n_3^3 n_o n_{\text{eff}}\lambda_{\text{p}}^6}\right) .
\label{quadm}
\end{equation}

The opposite limit (very large $|\eta|$) doesn't work, because $H_{\text{oe}}$ tends to zero for increasing $|\mathbf{a}|$. As a result, there are always regions where the argument of the cosh-function is small, even when $|\eta|$ is very large.

An alternative is to caclulate the expression order-by-order, using the expansion
\begin{align}
\ln_{\diamond}\left[\cosh_{\diamond}\left(4|H_{\text{oe}}|\right)\right]
\approx & \tfrac{1}{2}Z - \tfrac{1}{12} Z^{\diamond 2} + \tfrac{1}{45} Z^{\diamond 3} \nonumber \\
& - \tfrac{17}{2520} Z^{\diamond 4} + ...\ ,
\end{align}
where $Z \equiv 16 |H_{\text{oe}}|\diamond|H_{\text{eo}}|$. However, it still assumes that the kernel is relatively small. For the determinant, we need to compute the traces of all terms. The trace of the first term is given in Eq.~(\ref{trh2a}). For the higher-order terms, the integrals are intractable, but they can be simplified by applying suitable approximations.

In most experimental setups, the Rayleigh range of the pump beam is much longer than the thickness of the crystal. Hence, $\beta\ll 1$, which is the {\em thin-crystal approximation}. The phase-matching function (sinc-function) is then approximately equal to 1.

It is tempting to remove the sinc-function from the integral and evaluate the integral for the remaining expression. However, without the phase-matching function, the PDCS is not normalizable. So, the integral tends to diverge. A better approach is to replace the sinc-function by an exponential function with the negative argument of the sinc-function. However, this Gaussian approximation of the phase-matching function \cite{law} deviates from the sinc-function already at the sub-leading order in the expansion in terms of $\beta$ \cite{miatto3,noncol}. Its only purpose is to regularize the integral. Once the integral is evaluated, one sets $\beta\rightarrow 0$, except for an overall factor of $\beta$, which indicates that the result in the thin-crystal approximation is suppressed. Under the thin-crystal approximation, the PDC kernel in Eq.~(\ref{spdckern2}) is replaced by
\begin{align}
H_{\text{tc}}(\mathbf{a}_1,\mathbf{a}_2) = & \sqrt{2\pi} \eta\ \exp\left(-\tfrac{1}{2}\pi^2w_{\text{p}}^2\beta |\mathbf{a}_1-\mathbf{a}_2|^2\right) \nonumber \\
& \times \exp\left(-\pi^2 w_{\text{p}}^2|\nu_1\mathbf{a}_1+\nu_2\mathbf{a}_2|^2\right) .
\label{spdckerntc}
\end{align}
All the traces become tractable in the thin-crystal limit.

Another useful approximation is the {\em plane-wave approximation} where the pump beam radius is assumed to be much larger than any of the other parameters. As a result, one can treat the pump beam as a plane wave, leading to a two-dimensional Dirac delta function in the Fourier domain. The Dirac delta function is given by a limit process: when $w_{\text{p}}\rightarrow\infty$, the angular spectrum of the pump beam becomes
\begin{equation}
\lim_{w_{\text{p}}\rightarrow\infty} \exp\left(-\pi^2 w_{\text{p}}^2|\mathbf{a}_3|^2\right) = \frac{1}{\pi w_{\text{p}}^2} \delta(\mathbf{a}_3) ,
\end{equation}
where $\mathbf{a}_3 = \nu_1\mathbf{a}_1+\nu_2\mathbf{a}_2$. Using the plane-wave approximation, one can evaluate the integrals for the higher-order traces without modifying the sinc-functions.

The parameter dependences obtained under these two approximations are the same. The only differences are the numerical factors for the terms in the expansions. The traces of the even higher orders all have the form
\begin{equation}
\text{tr}\{Z^{\diamond n}\} = \xi A X^{2n} ,
\end{equation}
where $\xi$ is a numerical factor, and $A$ and $X$ are dimensionless quantities given by
\begin{align}
\begin{split}
A & = \frac{n_o^2 n_{\text{eff}}^2}{4 n_3^4\beta} , \\
X & = \frac{2\pi^2|\alpha_0|\sigma_{\text{II}} L}{n_o n_{\text{eff}} \lambda_{\text{p}}^2 w_{\text{p}}} \sqrt{\frac{\delta\lambda}{\lambda_{\text{p}}}} .
\end{split}
\label{axdefs}
\end{align}

\begin{figure}[th]
\includegraphics{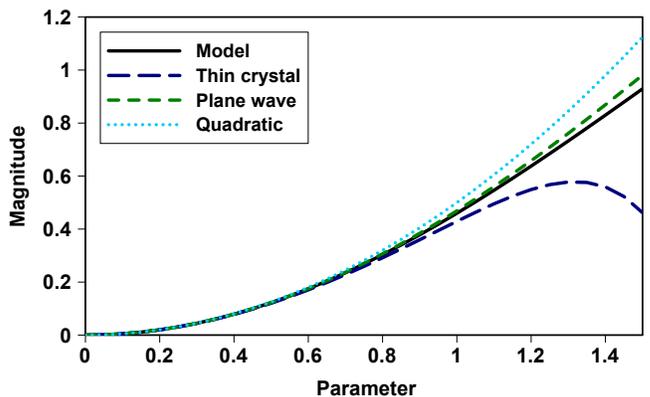}
\caption{Comparison of the curves for the model, the two approximations and the  small-amplitude model, plotted as functions of the dimensionless parameter $X$.}
\label{verg}
\end{figure}

Inspired by the expressions for these two approximations, we propose a closed-form expression for the full Schmidt number, given by
\begin{equation}
\mathcal{K} \approx \exp[A-A\cos(X)] ,
\label{volmodel}
\end{equation}
in terms of the quantities in Eq.~(\ref{axdefs}). We compare the curve for this model to those for the two approximations and the small-amplitude model for Eq.~(\ref{quadm}) in Fig.~\ref{verg}. Only the arguments of the exponentials are compared. Since these arguments share the same amplitude $A$, we remove it as well. So, the functions plotted in Fig.~\ref{verg} are
\begin{align}
\begin{split}
f_{\text{model}} & = 1-\cos(X) , \\
f_{\text{quad}} & = \tfrac{1}{2} X^2 , \\
f_{\text{tc}} & = \tfrac{1}{2} X^2 - \tfrac{1}{12} X^4 + \tfrac{16}{405} X^6 - \tfrac{17}{630} X^8 , \\
f_{\text{pw}} & = \tfrac{1}{2} X^2 - \tfrac{1}{9\pi} X^4 + \tfrac{11}{225\pi^2} X^6 - \tfrac{2567}{99225\pi^3} X^8 .
\end{split}
\label{modeldefs}
\end{align}
The comparison in Fig.~\ref{verg} shows that the curve for the model lies between those of the two approximations, while the quadratic curve lies above the others and thus over-estimates the magnitude for larger values of the parameter $X$. Based on the agreement among the different curves, we consider the model valid up to about $X\approx 0.8$.


In terms of the spatial degrees of freedom only, under the thin-crystal approximation, the expression for the Schmidt number is \cite{law}
\begin{equation}
\mathcal{K}_{\text{biphot}} \approx \frac{1}{2\beta} .
\end{equation}
Here it is assumed that the PDCS can be regarded as a biphoton state, and the thin-crystal approximation is implied by the use of the Gaussian approximation for the phase-matching function, which in turn implies that $\beta$ must be very small. An equivalent factor of $1/\beta$ appears in the definition of $A$ in Eq.~(\ref{axdefs}).

\section{\label{numexp}Numerical example}

To see how the incorporation of the particle-number degrees of freedom affects the Schmidt number, we'll consider an example with typical experimental parameters for such a parametric down-conversion experiment. Our hypothetical experiment, which is designed to satisfy the conditions for the thin-crystal and plane wave approximations, has a pump beam given by a 1~W pulsed laser with a pulse-repetition frequency of 100~MHz, a pulse width of 100~fs, a center wavelength of 400~nm and a waist radius in the crystal of 1~mm. The nonlinear crystal is a 10~mm long BBO crystal, cut for degenerate collinear down-conversion. Under these conditions, $\beta = 0.00211$, which gives $\mathcal{K}_{\text{biphot}} = 236.6$.

How does the incorporation of the particle-number degrees of freedom affects this result? There are $2.01\times 10^{10}$ photons per pulse for 1~W of optical power. It gives $|\alpha_0| = 0.14\times 10^6$. Here, the cross-section for the BBO is $\sigma_{\text{II}}=0.76\times 10^{-8}~\mu\text{m}^2$, so that $|\alpha_0|\sigma_{\text{II}} \approx 0.001~\mu\text{m}^2$ ($X=0.052$). The wavelength bandwidth is about 5~nm for a pulse width of 100~fs. The resulting full Schmidt number is only $\mathcal{K}_{\text{full}} = 1.15$.

To understand this number, we remind ourselves that the biphoton Schmidt number is conditioned on the detection of a biphoton. However, most pulses do not produce a down-converted biphoton under these conditions. The full Schmidt number, which is the average number of terms in the Schmidt decomposition, also considers the dominating vacuum state in those pulses without biphotons. In other words, the vacuum, which does not carry any spatiotemporal degrees of freedom, completely dominates the Schmidt decomposition.

\begin{figure}[th]
\includegraphics{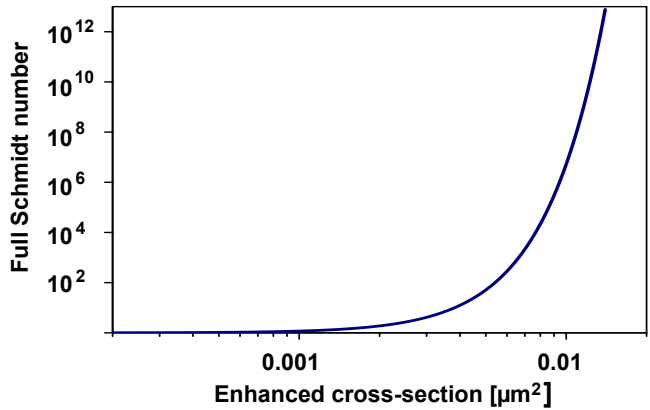}
\caption{Full Schmidt number plotted on logarithmic axes as a function of the enhanced cross-section $|\alpha_0|\sigma_{\text{II}}$.}
\label{volsn}
\end{figure}

One option to have more down-converted photons is to increase the power of the laser. For a 100~W laser the value of $|\alpha_0|$ increases by a factor of 10, leading to an enhanced cross-section of $|\alpha_0|\sigma_{\text{II}}\approx 0.01~\mu\text{m}^2$ ($X=0.52$). In this case, the full Schmidt number jumps to $\mathcal{K}_{\text{full}} = 2\times 10^6$. A logarithmic plot of the curve for the full Schmidt number, given in Eq.~(\ref{volmodel}), as a function of the enhanced cross-section $|\alpha_0|\sigma_{\text{II}}$ is shown in Fig.~\ref{volsn}. One can see a drastic increase of several orders of magnitude in the value of the full Schmidt number for one order of magnitude increase in the enhanced cross-section.

\section{\label{concl}Conclusions}

The main results of this paper are (a) the expression for the full Schmidt number of the PDCS that is obtain in Eq.~(\ref{hoof1}) with the aid of the Wigner functional formalism, without any severe restrictions on the spatiotemporal degrees of freedom and (b) the enormous values of the full Schmidt number shown in Fig.~\ref{volsn} --- much larger than previous estimates (see for example \cite{bsqvac}). While the latter is of special significance for any photonic quantum information system that uses a PDCS as a resource for high-dimensional entanglement, it is to our knowledge unprecedented in any field of quantum physics. The result suggests that measurements combining all the degrees of freedom may benefit significantly from the much larger entanglement dimension. It may for example produce significant improvements in quantum metrology, such as that which is to be applied in gravitational wave detection \cite{ligosqu}.

The final calculations to obtain a quantitative value for the full Schmidt number are still challenging. Fortunately, we could perform these calculations with the aid of some innocuous approximations. We did need to perform the calculation on an order-by-order basis, which limits how far we can plot the curves. Nevertheless, the result that we obtain at the 8-th order level can be plotted far enough to show the dramatic increase in the full Schmidt number. Beyond this point, it is reasonable to expect that the rise would eventually slow down, because the rate of increase in the full Schmidt number would eventually outpace the rate at which photons with a full complement of modes are added, based on an increase in the enhanced cross-section. To see the slow-down, a different approach is needed to evaluate the determinant. It is the hope that such an approach can be found in a future investigation.

\section*{Acknowledgement}

We acknowledge fruitful discussions with Thomas Konrad and Luis Lorenzo Sanchez Soto. This work is based on the research supported in part by the National Research Foundation (NRF) of South Africa (Grant Number: 118532).



\end{document}